\documentclass[paper]{ieice}

\usepackage{graphicx}
\usepackage{graphics}
\usepackage{latexsym}

\usepackage{algorithm}
\usepackage{algorithmic}

\newcommand{\tabincell}[2]{\begin{tabular}{@{}#1@{}}#2\end{tabular}}

\newcommand{\AmSLaTeX}{%
 $\mathcal A$\lower.4ex\hbox{$\!\mathcal M\!$}$\mathcal S$-\LaTeX}
\def\BibTeX{{\rmfamily B\kern-.05em
 \textsc{i\kern-.025em b}\kern-.08em
  T\kern-.1667em\lower.7ex\hbox{E}\kern-.125emX}}
 \makeatletter
\def\tmpcite#1{\@ifundefined{b@#1}{\textbf{?}}{\csname b@#1\endcsname}}%
\makeatother

 \field{A}
 \vol{92}
 \no{12}
 \SpecialIssue{}
 \SpecialSection{VLSI Design and CAD Algorithms}

\title{Voltage and Level-Shifter Assignment Driven Floorplanning}
\authorlist{%
 \breakauthorline{2}
 \authorentry[disyulei@gmail.com]
 {Bei Yu}{n}{labelA}
 \authorentry[dongsq@mail.tsinghua.edu.cn]
 {Sheqin Dong}{m}{labelA}
 \authorentry{Song Chen}{m}{labelB}
 \authorentry{Satoshi GOTO}{f}{labelB}
}
 \affiliate[labelA]{The authors are with the EDA lab, Department of Computer Science and Technology, Tsinghua University, Beijing 100084, China}
 \affiliate[labelB]{The authors are with the Graduate School of Information, Production and Systems, Waseda University, Kitakyushu-shi, 808-0135 Japan}

 \received{2009}{3}{18}
 \revised{2009}{6}{22}

\begin{document}
\maketitle

\begin{summary}
Low Power Design has become a significant requirement when the CMOS
technology entered the nanometer era. Multiple-Supply Voltage (MSV)
is a popular and effective method for both dynamic and static power
reduction while maintaining performance. Level shifters may cause
area and Interconnect Length Overhead(ILO), and should be considered
at both floorplanning and post-floorplanning stages. In this paper,
we propose a two phases algorithm framework, called VLSAF, to solve
voltage and level shifter assignment problem. At floorplanning
phase, we use a convex cost network flow algorithm to assign voltage
and a minimum cost flow algorithm to handle level-shifter
assignment. At post-floorplanning phase, a heuristic method is
adopted to redistribute white spaces and calculate the positions and
shapes of level shifters. The experimental results show VLSAF is
effective.
\end{summary}

\section{Introduction}

Low Power Design has become a significant requirement when the CMOS
technology entered the nanometer era. On the one hand, hundreds of
millions of transistors can be integrate on the same chip by using
system-on-chip(SoC) design methodologies. On the other hand, the
shrinking feature sizes and increasing circuit speed cause higher
power consumption, which not only shorten the battery life for
handheld devices but also lead to thermal and reliability problems.

Many techniques were introduced to deal with power optimization.
Among the existing techniques, MSV is a popular and effective method
for both dynamic and static power reduction while maintaining
performance. In the MSV design, one of the most important problem is
voltage assignment: timing critical modules are assigned to higher
voltage while noncritical modules are assigned to lower voltage, so
the power can be saved without degrading the overall circuit
performance.

\textit{Level-shifter} \cite{ICCAD02} has to be inserted to an
interconnect when a low voltage module drives a high voltage module
or a circuit may suffer from excessive short-circuit current and
leakage energy. From \cite{ICCAD06LEE} we can observe that the
number of level shifters increase rapidly as modules increase and
the area level-shifters consume can not be ignored. As a result,
level-shifters may cause area and performance overhead, and should
be considered during floorplanning and post-floorplanning stages.

There are a number of works addressing island generation and voltage
assignment in floorplanning and placement. Among these works,
voltage assignment is considered at various stages, including
pre-floorplanning\cite{ICCD05,ICCAD06LEE}; during
floorplanning\cite{ISLPED04,DAC08,ICCAD08MA}; and post-floorplaning
/ post-placement \cite{ASPDAC07,ICCAD07LEE,ICCAD05WU,ICCAD06ROY}.

Lee et al.\cite{ICCAD06LEE} handle voltage assignment by dynamic
programming, and  level shifters are inserted as soft block
according to the voltage assignment result at pre-floorplanning
stage. Then power network resource are considered during
floorplanning. However, there are some deficiencies in the work:
first, voltage assignment is handled before floorplanning, so
physical information such as the distances among modules are not
able to be taken into account; secondly, the search space is large
if level-shifters are considered as a module.

An approach based on ILP is used in \cite{ICCAD07LEE} for voltage
assignment at the post-floorplanning stage. Level-shifter planning
and power-network resources are considered. However, their approach
does not consider level-shifter's area consumption and relies on the
floorplanning result.

To make use of physical information such as the length of
interconnects among modules, voltage assignment problem should be
addressed during floorplanning. Ma et al.\cite{ICCAD08MA} transform
voltage assignment problem into a convex cost network flow problem,
and integrate it into floorplanning stage. However, their approach
consider neither level-shifters' area overhead nor level-shifters'
positions.


The remainder of this paper is organized as follows. Section 2
defines the voltage-island driven floorplanning problem. Section 3
presents our algorithm flow. Section 4 reports our experimental
results. At last, Section 5 concludes this paper.

\section{PROBLEM FORMULATION}

\newtheorem{define}{Definition}
\newtheorem{Theorem}{Theorem}
\newtheorem{problem}{Problem}
\newtheorem{lemma}{LEMMA}

In this paper, we use CBL\cite{CBL04} to represent every floorplan
generated. CBL is a topological representation dissecting the chip
into rectangular rooms, and each room is assigned at most one
module. Besides, all the nets are two-pin nets, and multi-pin nets
can be decomposed into a set of source-sink two-pin nets.
The wire length of every net is calculated by half-perimeter model.

\begin{define}[Interconnect Length Overhead]
Each \\level-shifter belongs to a net, we assume that a level
shifter can always be inserted in the net's bounding box. However,
if level-shifter is outside net's bounding box, its net's
interconnect length would increase. The increased length is denoted
as Interconnect Length Overhead (ILO).
\end{define}

\begin{define}[Power Network Resource] The power network
resource of a voltage island is evaluated by the half perimeter
wirelength of the minimal bounding box enclosing the island.
\end{define}

For every candidate floorplan, to meet the performance constraint,
timing-critical modules are assigned a high voltage, and the other
non-timing-critical modules are assigned a lower voltage to maximize
power saving. Besides, each level-shifter is assigned to a rough
position to minimize interconnect length overhead. We refer to the
problem as the Voltage and Level-Shifter Assignment driven
Floorplanning (VLSAF).

\begin{problem}{(VLSAF)}
We are given

\begin{flushleft}
\begin{tabular}{p{0.06cm}ll}
 1)&A set of m modules: $N=\{n_1, n_2, \dots, n_m\}$. Each \\
   &module $n_i$ is hard block(fixed size and aspect ratio),\\
   &and is given $k$ legal working voltages, and power\\
   &-delay tradeoff is represented as a delay-power curve \\
   &(DP--curve, as shown in Fig.\ref{E1}).\\
 2)&A netlist, which can be denoted as a directed acyclic\\
  &graph(DAG), $\hat G = (\hat V, \hat E)$, where $\hat V=\{ n_1, n_2,\dots,$\\
  &$n_m\}$, and $e(i,j) \in \hat E$ denotes an interconnect from\\
  &$n_i$ to $n_j$.\\
 3)&A timing constraint $T_{cycle}.$\\
 4)& Level-shifter's area, power and delay.\\
\end{tabular}
\end{flushleft}

After VLSAF, a chip floorplanning is generated to meet several
objectives: First, minimize the area and power cost. Secondly,
satisfying timing constraint. Third, insert all the level-shifters
in need and minimize the wire length and the interconnect length
overhead.
\end{problem}

\begin{figure}[tb]
\centering
\includegraphics[width=0.4\textwidth]{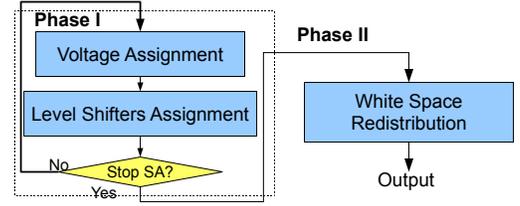}
\caption{~Overall of VLSAF} \label{overview}
\end{figure}

\section{VLSAF Algorithm}

\subsection{Overview of VLSAF}

As shown in Fig.\ref{overview}, algorithm VLSAF consists of two
phases: (I)voltage and level-shifters assignment during
floorplanning, (II) White Space Redistribution(WSR) at
post-floorplanning.

In Phase I, we modify the model in \cite{ICCAD08MA} to handle
voltage assignment and present a Min-Cost Max-Flow based method to
solve the level-shifters assignment problem. When generate a new
packing, we carry out voltage and level-shifter assignment. After
voltage assignment(VA), each module is assigned a voltage to reduce
power consumption as much as possible yet satisfies the performance
constraint. After level-shifter assignment(LSA), as many
level-shifters as possible are assigned a room. Level shifters which
can not assigned are belong to set $ELS$(detail in 3.4) and will
cause some Interconnect Length Overhead(ILO).

In Phase II, a heuristic method is adopted to calculate every
module's relative position in room. Besides, every room's white
space is divided into grids, and each level-shifter is decided its
aspect ratio and inserted to a grid. Finally, if a level-shifter can
not assign a room in LSA, it can be inserted into a room in order to
reduce interconnect length overhead(ILO).

\subsection{Voltage Assignment of Two Voltages}

During floorplanning, when a new floorplan is generated, we can
estimate the interconnect length between module i and module j,
denoted as $len_{ij}$. Similar to \cite{ICCAD08MA}, $len_{ij}$ can
be scaled to delay $delay_{ij}$ according to $delay_{ij} = \delta
\times len_{ij}$, where $\delta$ is a constant scaling factor. We
check every $delay_{ij}$, if $delay_{ij} \geq T_{cycle}$, then time
constraint can not be satisfied, so another new floorplan is
generated. Otherwise we carry out voltage assignment.

Given netlist $\hat G = (\hat V, \hat E)$, voltage assignment
problem can be formulated as (\ref{eq:0}):
\begin{equation}\label{eq:0}
    Minimize \sum_{i \in \hat V}P_i(d_i)
\end{equation}
\begin{displaymath}
    s.t. \left\{
      \begin{array}{lll}
          \mu_j-\mu_i \geq delay_{ij}+d_i  & \forall e(i,j)\in \hat{E}  & (1a)\\
          d_i\in \{d_{i}^1,d_{i}^2,\dots d_{i}^k \}       & \forall i \in \hat V       & (1b)\\
          0 \leq \mu_i \leq T_{cycle}      & \forall i \in \hat V       & (1c)\\
      \end{array}
    \right.
\end{displaymath}
where $\mu_i$ is the arrival time of vertex $i$ in DAG, and $d_i$ is
the delay of vertex $i$.

\subsubsection{Two Legal Working Voltages Assignment}

\begin{figure}[tb]
\centering
\includegraphics[width=0.5\textwidth]{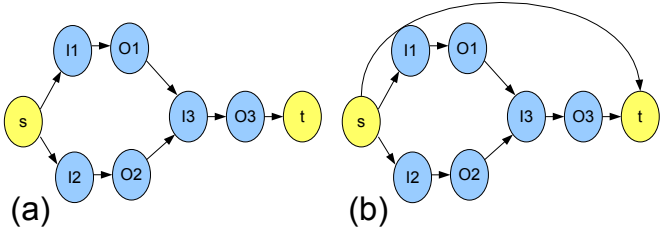}
\caption{~(a)$\bar G=\{\bar V,\bar E\}$, after adding nodes $s, t$
and diving nodes $N_i$ into $I_i$ and $O_i$ (b)Transformed $\bar
G=\{\bar V, \bar E\}$ by adding edge $e(s,t)$ to remove constraint
$\mu_t-\mu_s \leq T_{cycle}$ in equation (\ref{eq:1}).} \label{DAGs}
\end{figure}

When there are only two legal working voltages, we transform $\hat
G$ into $\bar G =(\bar V, \bar E)$. First, a start node $s$ and an
end node $t$ are added to $\hat V$, $s$ interconnect the nodes whose
in-degree are zero, and nodes with zero out-degree interconnect $t$.
We set $\bar V=\{s,t\}\cup \hat V=\{s, t, n_1, n_2, \dots , n_m\}$.
Besides, $n_i(i=1,\dots, m)$ are divided into two nodes: $I_i$ and
$O_i$, so $\bar V = \{s, t, I_1, O_1,  I_2, O_2, \dots, I_m, O_m
\}$. And $I_i$ is connected to $O_i$ by a directed edge. We denote
these new created edges $\{e(I_i,O_i)|I_i, O_i\in \bar V\}$ as $\bar
E_1$, denote edges $\{e(s,I_k)|I_k\in \bar V\}$ as $\bar E_3$, and
other edges as $\bar E_2$, and $\bar E=\bar E_1\cup \bar E_2\cup
\bar E_3$. The DAG $\bar G=(\bar V, \bar E)$ is shown in Fig.
\ref{DAGs} (a).


The mathematical program is in (\ref{eq:1}),where $d_{ij}$ is delay
from node i to node j.
\begin{equation}\label{eq:1}
    Minimize \sum_{e(i,j)\in{\bar{E}}}P_{ij}(d_{ij})
\end{equation}
\begin{displaymath}
    s.t. \left\{
      \begin{array}{lll}
          \mu_j-\mu_i \geq d_{ij}         & \forall e(i,j)\in \bar{E}   & (2a)\\
          \mu_t-\mu_s \leq T_{cycle}      &                             & (2b)\\
          d_{ij}\in \{d_{ij}^1,d_{ij}^2\} &\forall e(i,j)\in \bar E_1   & (2c)\\
          d_{ij}=delay_{ij}               & \forall e(i,j) \in \bar E_2 & (2d)\\
          d_{ij}=0                        & \forall e(i,j) \in \bar E_3 & (2e)\\
      \end{array}
    \right.
\end{displaymath}

\begin{figure}[tb]
\centering
\includegraphics[width=0.4\textwidth]{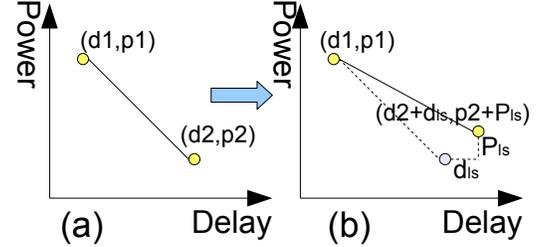}
\caption{~For a module, when number of legal working voltages is 2,
(a)original DP-curve, (b)modified DP-curve, adding the power and
delay of level-shifter.} \label{E1}
\end{figure}

Compare with \cite{ICCAD08MA}, which has more constraints as
follows:
\begin{displaymath}
    \left\{
      \begin{array}{ll}
          0 \leq \mu_i \leq T_{cycle}     & \forall i\in \bar V         \\
          l_{ij} \leq d_{ij} \leq u_{ij}  & \forall e(i,j)\in \bar E    \\
      \end{array}
    \right.
\end{displaymath}
we introduce some modifications. First, timing constraint used to be
estimated as $T_{cycle} - L \times d_{ls}$, where $L$ is the longest
path in DAG. To reduce tolerance of timing constraint, in module's
DP-curve, we add $d_{ls}$ to lower voltage's delay and add $p_{ls}$
to lower voltage's power(as shown in Fig. \ref{E1}), and time
constraint can be set as $T_{cycle}$. Since there are only two
possible supply voltages, power function $P_{ij}(d_{ij})$ still be
convex function. Secondly, we add start node $s$ and end node $t$ to
remove constraint $0 \leq t_i \leq T_{cycle}$. Third, since DP-curve
is a linear function, in other word, for $e(i,j)\in E_1$, $d_{ij}$
has only two choices: $d_{ij}^1$ and $d_{ij}^2$. We can prove later
that we can solve the program optimally even if we remove the
constraint $l_{ij} \leq d_{ij} \leq u_{ij}$.

We can incorporate constraints $(2b)$ and $(2a)$ by transforming
$(2b)$ into $\mu_s-\mu_t \geq -T_{cycle}$, and define $d_{st}$, s.t.
$\mu_t-\mu_s = d_{st}\quad \&\quad d_{st}\leq T_{cycle}$.
Accordingly, $\bar E_3=\{\bar E_3 \cup e(s,t)\}$, and the
transformed DAG $\bar G$ is shown in Fig.\ref{DAGs}(b). Besides, we
dualize the constraints $(2a)$ using a nonnegative Lagrangian
multiplier vector $\bar x$, obtaining the following Lagrangian
subproblem:
\begin{equation}\label{eq:4}
  L(\vec{x})=\textrm{min} \sum_{e(i,j)\in \bar E}[P_{ij}(d_{ij})+x_{ij}d_{ij}]+\sum_{i\in
  \bar V}x_{si}\mu_i
\end{equation}

We set $V=\bar V$, remove $e(i, j) \in E_3$, and add an edge $e(s,
i)$ for each node $i\in V$. The newly edges are denoted as $E_3$,
and $E_1=\bar E_1$, $E_2=\bar E_2$. Now $E=E_1 \cup E_2 \cup E_3$,
and the transformed DAG is denoted as $G=(V, E)$.

For every $e(s,i)\in E_3$, we set $d_{si}=\mu_i, P_{si}(d_{si})=0,
l_{si}=0$,$ u_{si}=\left\{
   \begin{array}{ll}
      K,         & if \quad i \ne t\\
      T_{cycle}, & if \quad i = t\\
   \end{array}
      \right.
$, where $k$ is a huge coefficient.

We define function $H_{ij}(x_{ij})$ for each $e(i,j)\in E$ as
follows:
$H_{ij}(x_{ij})=\textrm{min}_{dij}\{P_{ij}(d_{ij})+x_{ij}d_{ij}\} $.

For the $e(i,j)\in E_1$, because $P_{ij}(d_{ij})$ is linear function
\begin{equation}
P_{ij}(d_{ij})=-k\times d_{ij},\quad d_{ij}\in [d_{ij}^1, d_{ij}^2]
\end{equation}
where $k \geq 0$ and $-k$ denotes slope of the function,
$k=\frac{P_{ij}(d_{ij}^1)-P_{ij}(d_{ij}^2)}{d_{ij}^2-d_{ij}^1}$.

{\setlength\arraycolsep{2pt}
\begin{eqnarray}
H_{ij}(x_{ij}) & =& min\{(x_{ij}-k)\times d_{ij}\}{}\nonumber\\
               & =& \left \{
                      \begin{array}{ll}
                        (x_{ij}-k)\times d_{ij}^2 & 0 \leq x_{ij}\leq k\\
                        (x_{ij}-k)\times d_{ij}^1 & k \leq x_{ij}\\
                      \end{array}
                   \right.{}\nonumber\\
               & =& \left \{
                 \begin{array}{ll}
                   P_{ij}(d_{ij}^2)+d_{ij}^2x_{ij} & 0 \leq x_{ij}\leq k\\
                   P_{ij}(d_{ij}^1)+d_{ij}^1x_{ij} & k \leq x_{ij}\\
                 \end{array}
               \right.
\end{eqnarray}}

For the $e(i,j)\in E_2$, $H_{ij}(x_{ij})=d_{ij}x_{ij}, x_{ij}\geq
0$.

For the $e(i,j)\in E_3$,$ H_{ij}(x_{ij})=\left \{
    \begin{array}{ll}
      K_j \times x_{ij} & x_{ij}\leq 0\\
      0 & x_{ij}\geq 0\\
    \end{array}
  \right.$,
where $K_j = T_{cycle}$ if $j=t$; and if $j\ne t$, $K_j$ equals $K$.

To transform the problem into a minimum cost flow problem, we
construct an expanded network $G' = (V', E')$. There are three kinds
of edges to consider:
\begin{flushleft}
\begin{itemize}
\item
  $e(i,j)$ in E1:we introduce 2 edges in $G'$, and the costs of these
  edges are: $-d_{ij}^2, -d_{ij}^1$; upper capacities: $k,
  M-k$; lower capacities are both 0.
\item
  $e(i,j)$ in E2: cost, lower and upper capacity is $-d_{ij}$, 0, M.
\item
  Edge in E3: two edges are introduced in $G'$, one with cost, lower
  and upper capacity as ($-K_j, -M, 0$), another is ($0,0,M$).
\end{itemize}
\end{flushleft}

Using the cost-scaling algorithm, we can solve the minimum cost flow
problem in $G'$. For the given optimal flow $x^*$, we construct
residual network $G(x^*)$ and solve a shortest path problem to
determine shortest path distance $d(i)$ from node $s$ to every other
node. By implying that $\mu(i) = d(i)$ and $d_{ij} = \mu(i) -
\mu(j)$ for each $e(i,j) \in E_1$, we can finally solve voltage
assignment problem.

\subsubsection{Multi-Voltage Assignment}

When number of legal working voltages is more than two, we can solve
voltage assignment in a similar method.

\begin{define}[LS-DP-Curve]The power-delay tradeoff of level shifter is
represented by a LS-DP-Curve $\{(d_{ls}1,p_{ls}1),
(d_{ls}2,p_{ls}2), (d_{ls}3, p_{ls}3)\}$, where each pair $(d_{ls}i,
p_{ls}i)$ is the corresponding delay and power consumption when
level shifter is driving from module at voltage $i$.
\end{define}

When a module is at voltage $1$( the most high voltage ), it does
not need level shifter to drive other modules, $d_{ls}1=p_{ls}1=0$.
Lower voltage module needs bigger level shifter to drive other
modules. Since dynamic energy consumption is proportional to the
square of the supply voltage, it is trival that power increases
rapidly than delay. We assume the LS-DP-Curve is convex.

For each module, we modify its DP-Curve: replace each pair $(d_i,
p_i)$ by $(d_i+d_{ls}i, p_i+p_{ls}i)$, where $(d_{ls}i, p_{ls}i)$ is
level shifter's delay and power consumption.

\begin{lemma}
$f(x)$ is convex $\iff f(x_1+x_2)<\frac{f(x_1)+f(x_2)}{2}$, $\forall
x_1, x_2 \in Z$.
\end{lemma}
\begin{lemma}
If $f(x)$ and $g(x)$ are convex, then $P(x)=f(x)+g(x)$ is also
convex.
\end{lemma}

Using lemma 1 and lemma 2, we can prove that modified DP-Curve is
piecewise linear convex function with integer breakpoints, and we
can apply similar method like 3.2.1 to solve voltage assignment
problem.

\subsection{Level Shifters Assignment}

\begin{table}[bt]\label{table:notation}
\centering \caption{Notation used in LS Assignment}
\begin{tabular}{|l|l|}
 \hline $m$ & \# of modules \\
 \hline $n_{ls}$ & \# of level-shifters in need\\
 \hline $R$ & Set of rooms, $R = \{ r_1, r_2, \dots, r_m\}$\\
 \hline $r_j$    & Room containing module $j$\\
 \hline $ws_{j}$ & White space in $r_j$\\
 \hline $LS_{ij}$ & Set of LSs with same source $i$ and same sink $j$\\
 \hline $size_{ij}$ & \# of level shifters in $LS_{ij}$\\
 \hline $pws_j$ & Potential white space in room $r_j$\\
 \hline $w_{rj}(h_{rj})$ & Width(Height) of room $r_j$\\
 \hline $w_{mj}(h_{mj})$ & Width(Height) of module $n_j$\\
 \hline $w_{ij}(h_{ij})$ & Width(Height) of 1st Feasible Region $fr1_{ij}$\\
 \hline \hline
\end{tabular}
\end{table}

After voltage assignment, every module is assigned a voltage. Since
each net driving from a low voltage module to a high voltage module
should insert a level shifter, the number of level-shifters $n_{ls}$
is determined. To locate the $m$ modules, chip is dissected into set
of rooms $R = \{ r_1, r_2, \dots, r_m\}$. Due to the restriction
that level shifter cannot be placed on a module, the location must
be within a white space. Besides, level shifter has non-zero area,
it cannot be placed arbitrarily close to each other.

Here we carry out minimum cost flow based level-shifters assignment
to try to assign every level-shifters one room. We define sets of
level shifters $LS = \sum_{i=1}^n \sum_{j=1}^n LS_{ij}(i=1,\dots, n;
j=1,\dots, n; i \ne j)$, every set $LS_{ij}$ contain $size_{ij}$
level shifters with same source module $i$ and the same sink module
$j$, and $\sum_{i=1}^n \sum_{j=1}^n size_{ij} = n_{ls}$.

To check whether a room has extra space to insert level-shifter, we
denote the White Space in room $r_j$ as $ws_j$, whose area can be
calculated as follow:
\begin{equation}
    Area(ws_j) = w_{rj}\times h_{rj} - w_{mj}\times h_{mj} \\
\end{equation}
where $w_{rj}(h_{rj})$ denotes the width(height) of room $r_j$,
$w_{mj}(h_{mj})$ denotes the width(height) of module $n_j$.

\begin{figure}[tb]
\centering
\includegraphics[width=0.47\textwidth]{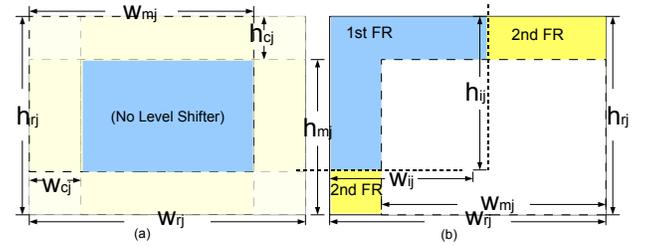}
\caption{~(a)No matter how to move the module, dark area can not
insert level-shifter, while blank area is Potential White Space(PWS)
of $R_j$ (b)1st and 2nd Feasible Region of $FR_{ij}$.}
\label{PWSandFR}
\end{figure}

Each level-shifter belongs to a net, and is inserted into white
space. If white space is outside the net's bounding box, inserting
level shifter may cause Interconnect Length Overhead(ILO), so each
white space has its own cost for given level shifter. Since we
assume all modules are hard blocks, some space of room must belong
to a module(as shown in Fig.\ref{PWSandFR}(a), center dashed area
can not insert level shifter no matter how to put the module).

\begin{define}[Potential White Space (PWS)]
The space of room $r_j$ can be white space through module moving is
denoted as Potential White Space($pws_{j}$).
\end{define}

$pws_j$ can be considered as two horizontal channels and two
vertical channels, as shown in Fig. \ref{PWSandFR} (a), we denote
the width of vertical channel as $w_{cj}=w_{rj}-w_{mj}$, and the
width of horizontal channel as $h_{cj}= h_{rj} - h_{mj}$.

\begin{define}[Feasible Region (FR)]
For a net requiring level shifter $i$, its bounding box is denoted
as $b_i$, we define level-shifter $i$'s feasible region as $FR_i$
and $FR_i = \{ ws_j | \forall j, b_i \cap r_j \ne 0 \}$.
\end{define}

For room $r_j$, if its white space $ws_j$ belongs to level-shifter
$ls_i$'s Feasible Region $FR_i$, we call $r_j$ as $ls_i$'s candidate
room. The part of $ws_j$ in $b_i$ is denoted as 1st Feasible
Region($fr1_{ij}$), while the other part of $ws_j$ is denoted as 2nd
Feasible Region($fr2_{ij}$). If $ls_i$ is inserted into its
candidate room, then will not cause Interconnect Length Overhead
(ILO) to its net.

We set $w = max(w_{ij}-w_{cj},0)$ and $h = max(h_{ij}-h_{cj},0)$,
then the area of $fr1_{ij}$ can be calculated as follows:
\begin{equation}
    Area(fr1_{ij})=w_{ij}\times h_{ij} - w \times h
\end{equation}

\begin{figure}[tb]
\centering
\includegraphics[width=0.4\textwidth]{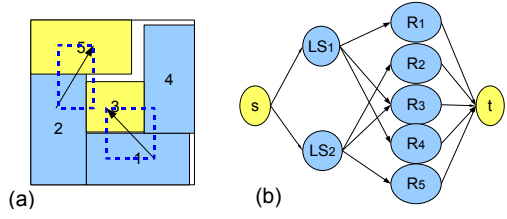}
\caption{~(a)$LS_{13}$ drives from module 1 to module 3 and
$LS_{25}$ drives from module 2 to module 5. (b)Corresponding network
graph, $LS_{1}$ can be assigned to room 1,3,4, while $LS_{2}$ can be
assigned to room 2,3,5.} \label{network}
\end{figure}

We construct a network graph $G^* = (V^*, E^*)$, and then use a
min-cost max-flow algorithm to determine which room each level
shifter belong to. If all level shifters are assigned to their
candidate rooms, no ILO will occur. A simple example is shown in
Fig.\ref{network}.

\begin{flushleft}
\begin{itemize}
\item
  $V^* = \{s, t\}\cup LS \cup R$.
\item
  $E^* = \{ (s, ls_i)|LS_i\in LS\} \cup \{ ( LS_i, r_j) | \forall r_j\ is\ LS_i 's\ candidate\ room \} \cup \{( r_j, t)| r_j \in R \}$.
\item
  Capacities: $C(s,LS_i)=size_i, C(LS_i, r_j)=size_i, C(r_j,
  t ) = \frac {Area(ws_j)} {a_{ls}}$.
\item
  Cost: $F(s, LS_i)=0, F(r_j, t)=0; F(LS_i, r_j)=F_{ij}$, which will
  discussed below.
\end{itemize}
\end{flushleft}

We define area percent of $fr1_{ij}$ as $p_{ij}$, $0\le p_{ij} \leq
1$.
\begin{equation}
  p_{ij} =\left \{
    \begin{array}{ll}
  \frac{Area(fr1_{ij})} {Area(ws_{j})}, &
  Area(ws_j)\ne 0\\
  0, & others\\
    \end{array}
  \right.
\end{equation}

Define cost of edge $e(LS_i, r_j)$, $F_{ij}$ is a function of
$p_{ij}$:
\begin{eqnarray}\label{eq:fij}
  F_{ij}(p_{ij})&=&\lceil \frac{1}{p_{ij}+\mu} + (1-p_{ij}) \times k
  {}
  \nonumber\\
    &&{}\times (Term1_{ij} + Term2_{ij} ) \rceil
\end{eqnarray}
where $\mu$ is a small coefficient, $k$ is a undetermined
coefficient and $Term1_{ij}, Term2_{ij}$ is penalty terms, and
$Term1_{ij} = \left \{
    \begin{array}{ll}
      \frac {h_{rj} - h{ij}}{w_{cj}}, & w_{cj}\ne 0\\
      0,                              & w_{cj} = 0\\
    \end{array}
  \right.$, $Term2_{ij} = \left \{
    \begin{array}{ll}
      \frac{w_{rj}-w_{ij}}{h_{cj}}, & h_{cj}\ne 0\\
      0,                            & h_{cj} = 0\\
    \end{array}
  \right.$.

Equation (\ref{eq:fij}) has some special characters. First, it is a
monotonically decreasing function of $p_{ij}$, which means we are
inclined to put level-shifter in the room which has higher
percentage of 1st $FR$. Besides, it can not be too large even
$fr1_{ij}$ is very small, so we add coefficient $\mu$ and $max
F_{ij}(p_{ij}) \simeq \lceil \frac{1}{\mu}\rceil$. Third, we observe
that even two room have the same $p_{ij}$ and $p_{ij} \le 1$, if
level shifter is inserted in $fr2_{ij}$, the room has longer
$fr2_{ij}$ may cause longer length. Consequently, in equation
(\ref{eq:fij}), we add the penalty term $Term1_{ij}$ and
$Term2_{ij}$.

It can be shown that any flow in the network $G^*$ assigns level
shifters to white spaces (given by the saturated edges between the
level shifters $LS_i$'s and the white space nodes $ws_j$'s).
Although level shifter assignment is similar to buffer assignment,
each net has at most one level shifter to insert and it can be
solved effectively by minimum cost flow algorithm(run in polynomial
time\cite{book:flow}).

\subsection{White Space Redistribution (WSR)}

During floorplanning, voltage assignment and level shifter
assignment are carried out for each candidate solution. Best
solution that satisfies constraints and inserts most level shifters
would be stored. After floorplanning, most level-shifters can be
assigned to rooms in stored best solution. We define $ELS$ a set
which contains level-shifters that can not be assigned to any room.
In room $r_j$, we define the module to pack as $n_j$, and a group of
level shifters to insert as $Ls_j=\{ls_1, ls_2, \dots, ls_{pi}\}$.
Follow condition must be satisfied:
\[Area(n_j) +\sum_{k=1}^{pi}Area(ls_k) \leq Area(r_j)\]

Traditional room-based floorplanner will pack the modules at the
lower-left corner or the center of the rooms. Different from the
traditional block planning method, to favor the level-shifters
insertion, a heuristic method( called WSR) is adopted to calculate
modules' and level-shifters' relative positions in rooms. The
framework of algorithm WSR is shown in Algorithm \ref{alg:WSR}.

\begin{algorithm}[tb]
 \caption{(WSR)}
 \label{alg:WSR}
\begin{algorithmic}[1]
 \FOR{$j=1$ to $m$}
   \STATE $pj \leftarrow $ sizeof($Ls_j$);
   \STATE $F_{right}\leftarrow 0, F_{left}\leftarrow 0, F_{up}\leftarrow 0, F_{down}\leftarrow 0$;
   \FOR{$i=1$ to $pj$}
     \STATE calculate $F_{ix}$ and $F_{iy}$;
     \STATE update $F_{right}, F_{left}, F_{up}, F_{down}$;
   \ENDFOR
   \STATE calculate $X_{n}$ and $Y_{nj}$;      /*Relative Position*/
   \STATE generate grids $G_j$ in white space;
   \STATE sort $Ls_j$ by priority;
   \FOR{$i=1$ to $pj$}
     \STATE pick one grid to insert $ls_i$; /*Level shifter insertion*/
   \ENDFOR
 \ENDFOR
 \STATE InsertELS();
 \FOR{$j=1$ to $m$}
   \STATE move modules $n_j$ under demand of Power Network;
 \ENDFOR
\end{algorithmic}
\end{algorithm}

\subsubsection{Relative Position Calculation}

\begin{figure}[tb]
\centering
\includegraphics[width=0.2\textwidth]{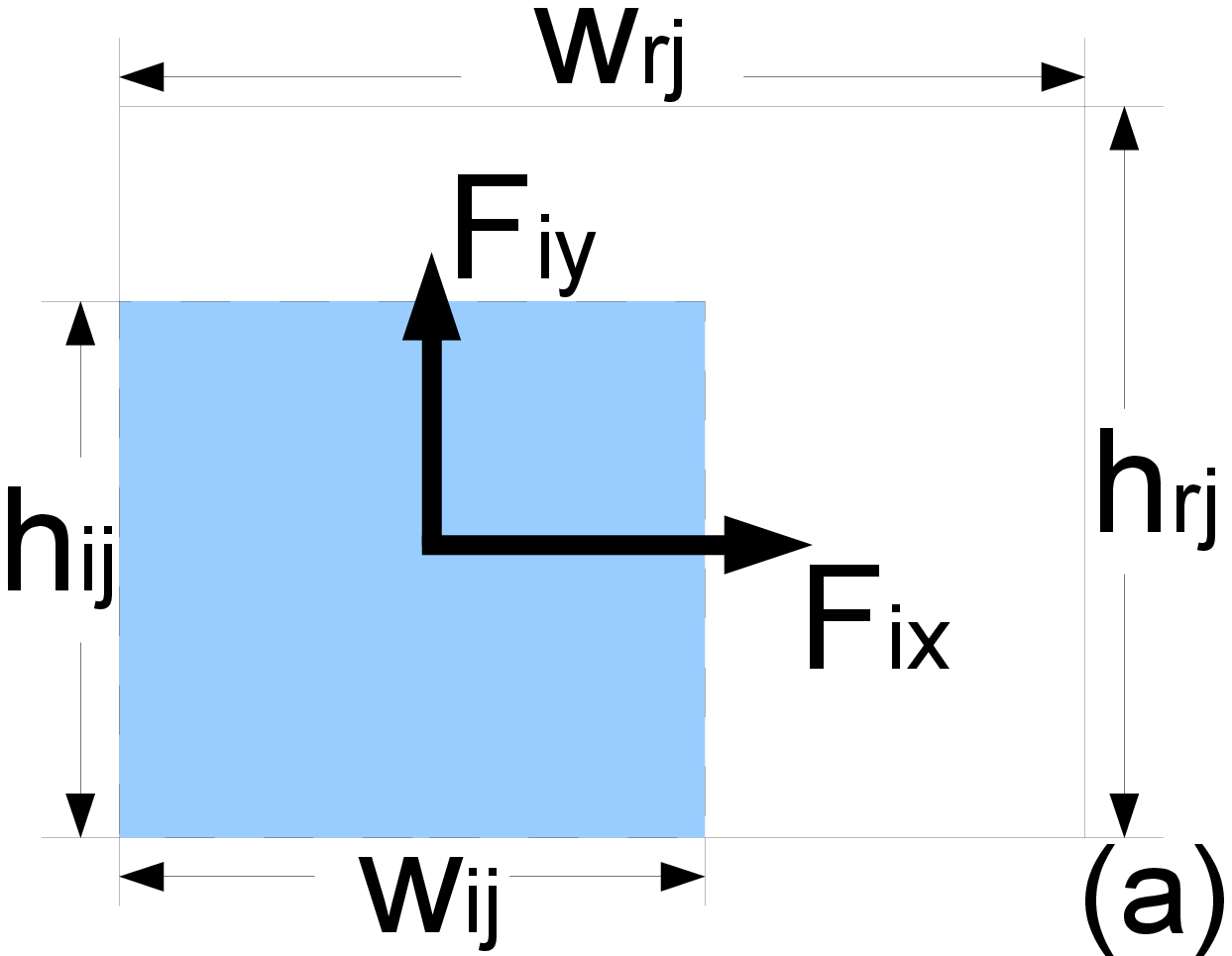}
\includegraphics[width=0.22\textwidth]{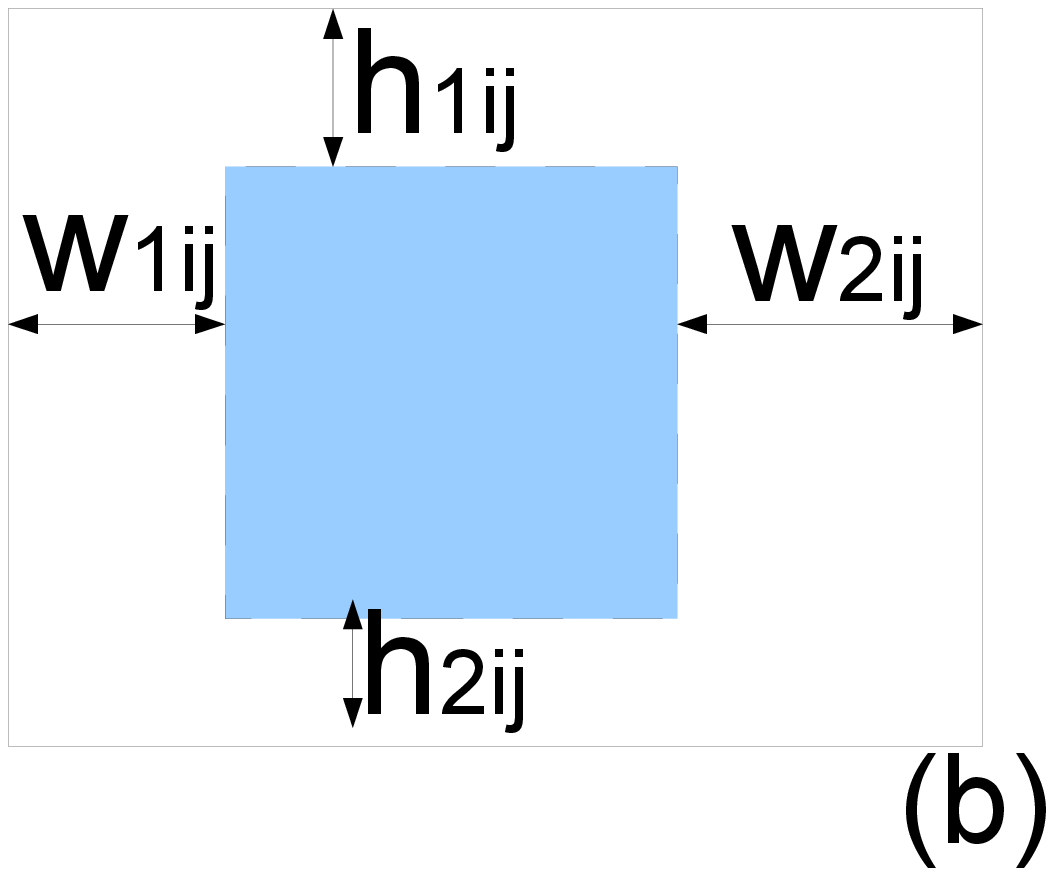}
\caption{~In room $r_j$, (a)if level-shifter $ls_i$ prefers to
locate in lower-left corner (dark area is prefer region), then
$ls_i$ produces forces $(F_{ix}, F_{iy})$ to pushes module $n_j$
upper and right. (b)$w_{1ij}, w_{2ij}, h_{1ij}, h_{2ij}$ are defined
to calculate forces $(F_{ix}, F_{iy})$.}\label{Tk}
\end{figure}

If a level-shifter $ls_i$ is assigned into room $r_j$, a prefer
region is provided. If $ls_i$ is inserted in the prefer region, then
interconnect would not lengthen. For each level-shifter to insert in
room $r_j$, a force is produced to push the module $n_j$ apart from
the level-shifter. We consider the force produced by $ls_i$ in x-
and y-direction separately, denoted as $F_{ix}$ and $F_{iy}$. For
example, as shown in Fig. \ref{Tk}(a), if $ls_i$ prefers to locate
in the lower-left corner of $r_j$, then $F_{ix}$ pushes $n_j$ to
right and $F_{iy}$ pushes $n_j$ to upper. To calculate $F_{ix}$ and
$F_{iy}$, prefer area is defined as a quaternion $(w_{1ij}, w_{2ij},
h_{1ij}, h_{2ij})$, where $w_{1ij}(w_{2ij})$ is the distance from
prefer area to left(right) boundary of $r_j$, $h_{1ij}(h_{2ij})$ is
the distance from prefer area to upper(lower) boundary of $r_j$, as
shown in Fig. \ref{Tk}(b).

$F_{ix}$ and $F_{iy}$ can be calculated as equation (\ref{eq:fjx}).
\begin{equation}\label{eq:fjx}
  \begin{array}{ll}
      F_{ix} = \frac{ w_{2ij}-w_{1ij} }{ w_{rj} },  &
      F_{iy} = \frac{ h_{2ij}-h_{1ij} }{ h_{rj} }\\
  \end{array}
\end{equation}

To calculate the position of module $n_j$, we define four variables
$F_{right}, F_{left}, F_{up}, F_{down}$ as follows:
\begin{equation}
  \left \{
    \begin{array}{lll}
      F_{right} &= \sum_{i} F_{ix}, & \forall F_{ix} \geq 0\\
      F_{left}  &= \sum_{i} F_{ix}, & \forall F_{ix} < 0\\
      F_{up}    &= \sum_{i} F_{iy}, & \forall F_{iy} \geq 0\\
      F_{down}  &= \sum_{i} F_{iy}, & \forall F_{iy} < 0\\
    \end{array}
  \right.
\end{equation}

Relative position of $n_j$ in room $r_j$ is denoted as $(X_{nj},
Y_{nj})$, then $X_{nj} = \frac { (w_{rj}-w_{mj})\times F_{right} } {
F_{right} - F_{left} }$ and $Y_{nj} = \frac { (h_{rj}-h_{mj})\times
F_{up} } { F_{up} - F_{down} }$.

\begin{figure}[tb]
\centering
\includegraphics[width=0.2\textwidth]{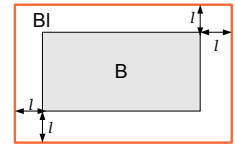}
\caption{~$B_l$ is $l$-bounding box of $B$.} \label{lbox}
\end{figure}

\subsubsection{Grids Generation and LS Insertion}

\begin{define}[$l$-bounding box]Given a level shifter $ls_k$, we define the bounding box of $ls_k$'s net as $B$, whose
width is $wid_B$ and height is $hei_B$. The $l$-bounding box of $B$
is $B_l$, which has the same centric position. Besides, width of
$B_l$ is $(wid_B+2 \times l)$ and height is $(hei_B+2 \times l)$ (as
shown in Fig.\ref{lbox}).
\end{define}

In room $r_j$, after calculating module $n_j$'s relative position,
at most four rectangular white spaces are generated. We divide each
white spaces into rectangular grids, whose area are all $a_{ls}$. So
room $r_j$ records a set of grids $G_j=\{ g_1, g_2, \dots, g_{m}, m
\times a_{ls} \leq Area(r_j) - Area(n_j)\}$, and each grid has its
position. Level-shifters in set $Ls_j$ are sorted by area of prefer
region. Smaller prefer region, higher priority. Then each
level-shifter picks one grid in order.

After every level-shifter assigned choosing a grid, each level
shifter $ls_k$ in ELS chooses one free grid to insert(as shown in
Algorithm \ref{alg:els}).

\begin{algorithm}[htb]
 \caption{InsertELS()}
 \label{alg:els}
\begin{algorithmic}[1]
 \STATE Initialize $l=0$, $step$;
 \WHILE{$ELS$ is not empty}
   \STATE $l \leftarrow l+step$;
   \STATE $num \leftarrow ELS.size()$;
   \FOR{$i=1$ to $num$}
     \STATE Generate $l$-bounding box of $ls_i$;
     \STATE Find all free grids inside $l$-bounding box;
   \ENDFOR
   \STATE Construct bipartite graphs;
   \STATE Solve maximum bipartite matching;
   \STATE Update $ELS$;
 \ENDWHILE
\end{algorithmic}
\end{algorithm}

\begin{table*}[tb]\label{table:result}
\centering \caption{The Comparison Between the VLSAF and the
Previous Work }
\begin{tabular}{|c|c|c|c|c|c|c|c|c|c|c|c|}
\hline \hline Benchmark&Max Power&\multicolumn{2}{|c}{Power
Cost}&\multicolumn{2}{|c}{PNR}&\multicolumn{2}{|c}{LS
Number}&\multicolumn{2}{|c}{W.S(\%)}&\multicolumn{2}{|c|}{Time(s)}\\
\cline{3-12} & &
\cite{ICCAD06LEE}&VLSAF&\cite{ICCAD06LEE}&VLSAF&\cite{ICCAD06LEE}&VLSAF&\cite{ICCAD06LEE}&VLSAF&\cite{ICCAD06LEE}&VLSAF
\\ \hline
 n10 &216841& 216840&189142& 965&1007& 0&9&4.87&9.44&6.001&3.24\\
\hline n30 &205650&190717&146483&1369&1436&57&25&9.03&11.32&115.07&35.11\\
\hline
n50&195140&172884&135316&1514&1460&119&114&21.10&16.66&569.36&116.97\\
\hline
n100 &180022&179876&123526&1671&1354&92&153&34.07&26.71&1768&688.13\\
\hline n200 &177633&174818&130050&2040&1763&399&203&46.52&29.66&4212&1969.12\\
\hline n300 &273499&219492&234389&2147&1997&452&337&44.10&37.74&4800&2392.8\\
\hline Avg & -&192438&159818&1617.7&1502.8&186&140.2&26.61&21.92&1911.74&857.56\\
\hline Diff & -&-&-17\%&-&-7.2\%&-&-24.7\%&-&-17.6\%&-&-55.2\%\\
\hline \hline
\end{tabular}
\end{table*}

\begin{table*}[tb]\label{table:compare}
\centering \caption{VLSAF v.s. VAF+LSI}
\begin{tabular}{|c|c|c|c|c|c|c|c|c|}
\hline \hline & \multicolumn{2}{|c}{Wire Length w. LS} &
\multicolumn{2}{|c}{ILO(\%)}&\multicolumn{2}{|c|}{W.S(\%)}&\multicolumn{2}{|c|}{Time(s)} \\
\cline{2-9} & VLSAF& VAF+LSI & VLSAF& VAF+LSI & VLSAF& VAF+LSI & VLSAF& VAF+LSI\\
\hline n10&13552&17937&0.89&2.29&9.44&10.46&3.24&2.09 \\
\hline n30&44225&43282&0.31&0.85&11.32&10.75&35.11&23.13 \\
\hline n50&92678&95666&1.20&2.27&16.66&18.12&116.97&39.81 \\
\hline n100&185622&191522&1.03&2.40&26.71&26.40&688.13&327.01 \\
\hline n200&366003&365792&1.64&4.28&29.66&30.06&1969.12&1304.3 \\
\hline n300&560042&600348&0.67&1.37&37.74&35.36&2392.8&1772.03 \\
\hline Avg&210404&219091&0.96&2.24&21.92&21.86&857.56&578.06\\
\hline Diff&-&+4\%&-&+133\%&-&-0.3\%&-&-32.5\%\\ \hline \hline
\end{tabular}
\end{table*}

\begin{table*}[tb]\label{table:result2}
 \centering
 \caption{Experimental Results with More Legal Working Voltage }
 \footnotesize
 \begin{tabular}{|c|c|c|c|c|c|c|c||c|c|c|c|c|c|c|c|}
   \hline \hline
    & $k$ & \tabincell{c}{Power\\Cost} & \tabincell{c}{Wire\\Length} & \tabincell{c}{LS\\Num} &
   \tabincell{c}{ILO\\(\%)} & \tabincell{c}{W.S\\(\%)} &
   \tabincell{c}{Time\\(s)} &
    & $k$ & \tabincell{c}{Power\\Cost} & \tabincell{c}{Wire\\Length} & \tabincell{c}{LS\\Num} &
   \tabincell{c}{ILO\\(\%)} & \tabincell{c}{W.S\\(\%)} &
   \tabincell{c}{Time\\(s)} \\
   \hline
   n10  & 3 & 163352 & 16386  & 10  & 0.13 & 11.58 & 3.03 &
   n100 & 3 & 131394 & 180023 & 150 & 0.50 & 26.8  & 438.05 \\
        & 4 & 162794 & 16474  & 11  & 0.12 & 11.54 & 3.96 &
        & 4 & 120885 & 181280 & 167 & 0.34 & 26.07 & 414.7 \\
   \hline
   n30  & 3 & 139466 & 45103 & 42 & 0.32 & 15.85 & 20.82 &
   n200 & 3 & 112801 & 331627 & 242 & 0.55 & 35.44 & 1949.4\\
        & 4 & 138463 & 45388 & 42   & 0.21 & 17.63 & 19.83 &
        & 4 & 117538 & 344111 & 248 & 0.46 & 34.84 & 2036.4 \\
   \hline
   n50  & 3 & 132199 & 94105  & 130 & 0.37 & 22.72 & 51.10 &
   n300 & 3 & 218636 & 556718 & 389 & 0.44 & 37.14 & 2390.2\\
        & 4 & 133564 & 93296  & 151 & 0.50 & 22.95 & 49.35 &
        & 4 & 206354 & 568364 & 417 & 0.53 & 38.54 & 2377.2\\
   \hline \hline
 \end{tabular}
\end{table*}

Given $l$, for each level shifter $ls_k$ in $ELS$, we construct a
$l$-bounding box, called $B_l^k$ (step 6). Then we find all free
grids in $B_l^k$ (step 7). In step 9, we construct bipartite graphs,
then we use Hungarian algorithm to find maximum bipartite matching,
which takes O(mn) time\footnote{$m$ is the number of edges, and $n$
is the number of nodes}(step 10). In step 11, we update $ELS$, and
remove level shifters that have been inserted. If there still are
level shifters in $ELS$, we update $l$ and go back to step 6.

After InsertELS(), in room $r_j$, if not all the grids are inserted
by level shifter, module $n_j$ may remove. If $n_j$ is in lowest
voltage, it removes toward left and down to reduce total area.
Otherwise, it removes toward the center of power network to minimize
power network resource.


\section{EXPERIMENTAL RESULTS}

\begin{figure}[tb]
\centering
\includegraphics[width=0.23\textwidth]{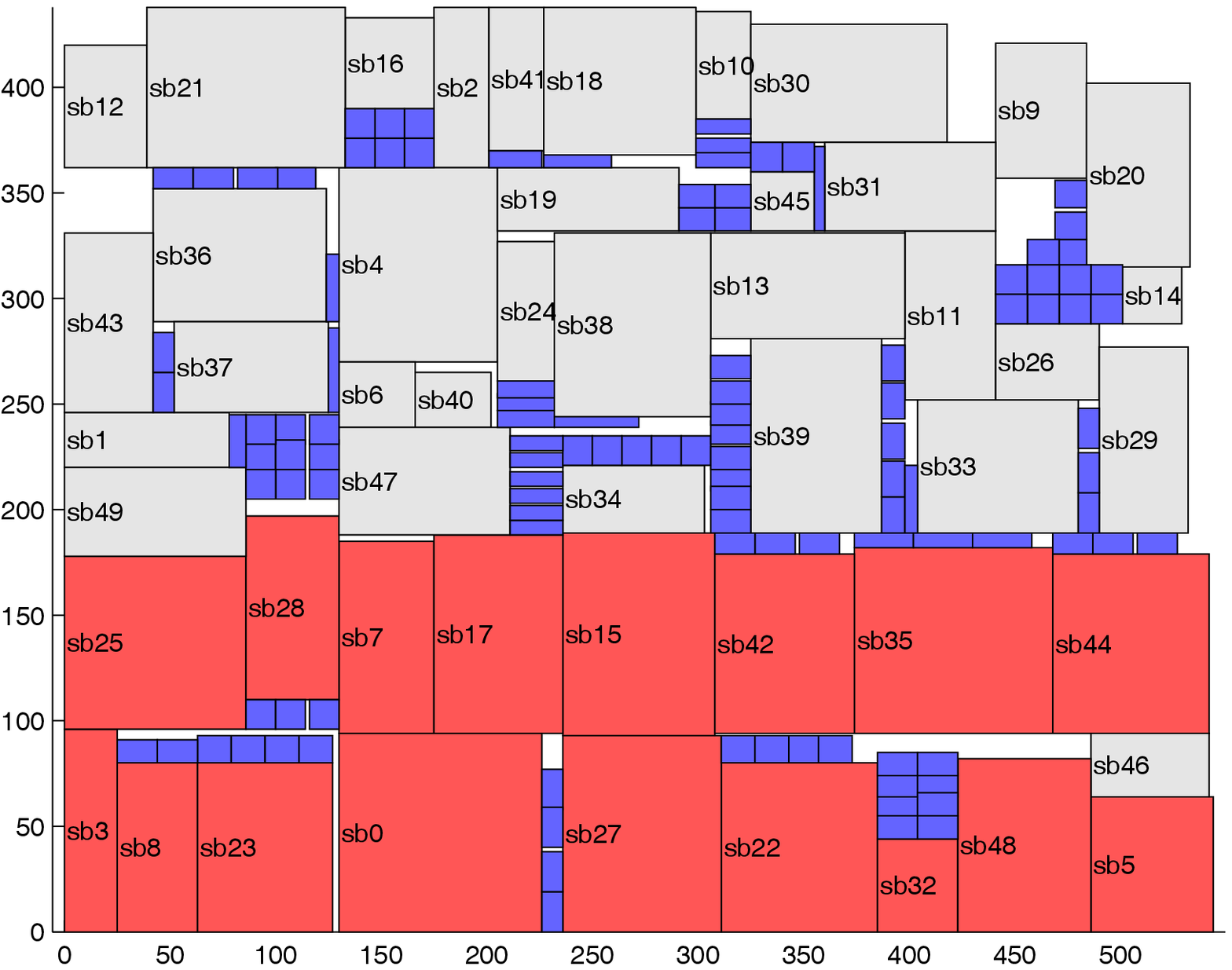}
\includegraphics[width=0.23\textwidth]{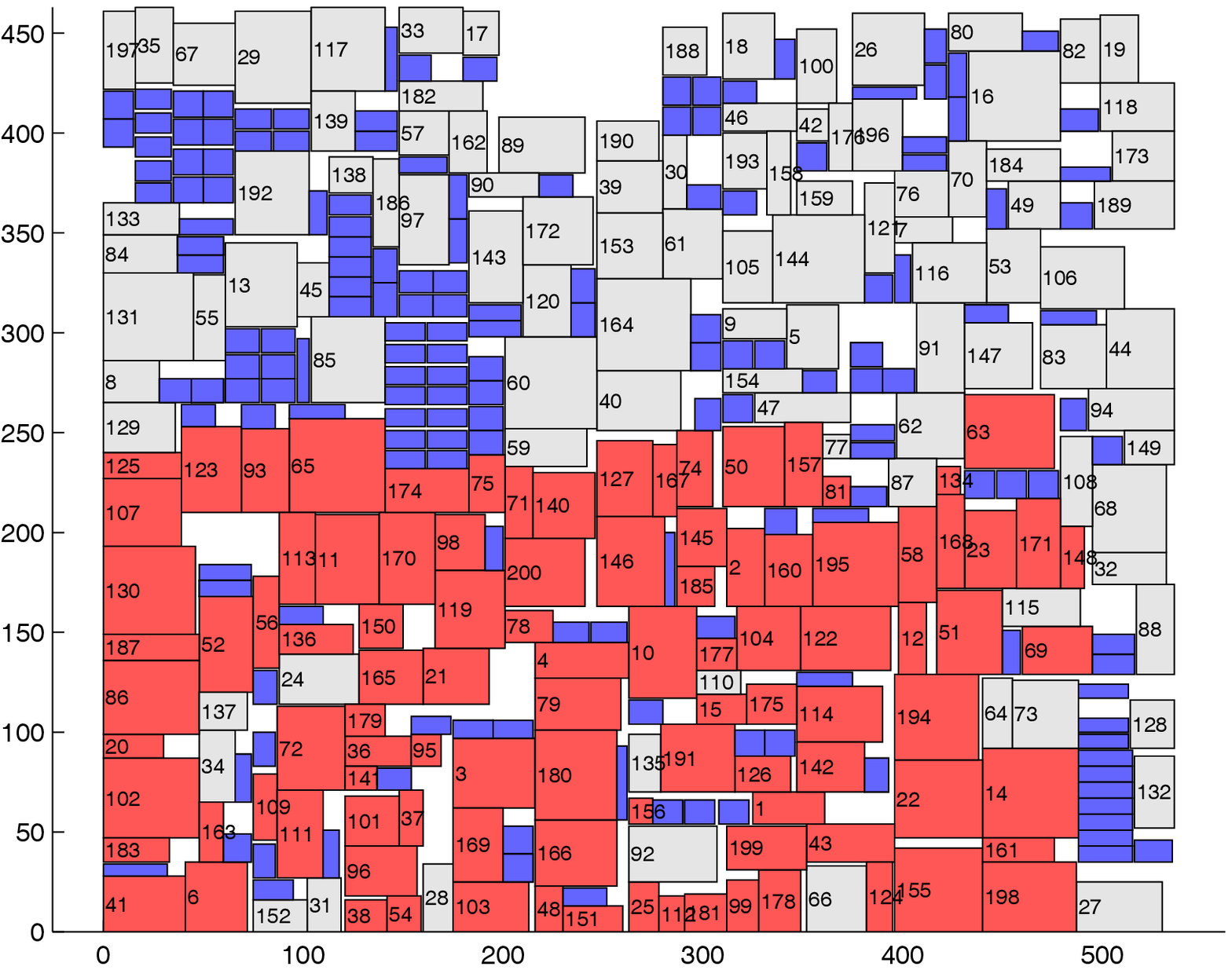}
\caption{Experimental results of n50 and n200 with two legal working
voltages.}\label{result}
\end{figure}

We implemented algorithm VLSAF in the C++ programming language and
executed on a Linux machine with a 3.0GHz CPU and 1GB Memory. Fig.
\ref{result} shows the experimental results of the benchmarks n50
and n200. Blocks in the same voltage are nearly clustered together
to reduce the power-network resource, and level shifters (small dark
blocks) are inserted in white spaces. Cost function in simulated
annealing is:
\begin{displaymath}
\Phi = \lambda_AA + \lambda_WW + \lambda_PP + \lambda_RR +
\lambda_NN
\end{displaymath}
where $A$ and $W$ represent the floorplan area and wire length; $P$
represents the total power consumption; $R$ represents the power
network resource; and $N$ records the number of level shifters that
can not be assigned.

The previous work \cite{ICCAD06LEE} is the recent one in handling
floorplanning problem considering voltage assignment and
level-shifter insertion. To compare with \cite{ICCAD06LEE}, we
performed our experiments on the same test cases, which are based on
the GSRC benchmarks adding power and delay specifications. Table $1$
shows comparisons between our experimental result and
\cite{ICCAD06LEE}. The column Power Cost means the actual power
consumption, column PNR means power network resource consumption and
the column W.S means white space. VLSAF can save 17\% power and
7.2\% PNR. The White Space and Run Time results show our framework
is about 2X faster while white space can be saved by 17.6\%.

We further demonstrated the effectiveness of our approach by
performing another contrastive approach VAF+LSI, which solves
level-shifter assignment and insertion only at post-floorplan stage.
Table $3$ compares VLSAF and VAF+LSI. We can see that in VAF+LSI,
although runtime is shorter(no iterative level shifter assignments
during floorplanning), wire length and interconnect length
overhead(ILO) are increased by 4\% and 133\%. High ILO may cause
delay estimation among modules inaccurate, or even lead to timing
constraint violation. Accordingly, VLSAF is effective and
significant with a reasonable more runtime.

Besides, we have done two sets of experiments in which the number of
legal working voltages for each module is set three and four. The
detailed results are listed in Table 4.

\section{CONCLUSIONS}

We have proposed a two phases framework to solve voltage assignment
and level shifter insertion: phase one is voltage and level-shifter
assignment driven floorplanning; phase two is white space
redistribution at post- floorplanning stage. Experimental results
have shown that our framework is effective in reducing power cost
while considering level shifters' positions and areas.


\profile{Bei Yu}{ received the B.E degree in the Department of
Mathematic from UESTC, China in 2007. He is currently a M.E.
candidate in EDA lab, Department of Computer Science and Technology,
Tsinghua University, China. His research interests include CAD for
VLSI, floorplanning algorithms and low power design.}

\profile{Sheqin Dong}{ received the B.E. degree in Computer Science
in 1985, M.S. degree in semiconductor physics and device in 1988,
and Ph.D. degree in mechantronic control and automation in 1996. He
is currently an associate professor of the EDA lab at the department
of computer science and technology in Tsinghua University. His
current research interests include CAD for VLSI, floorplanning and
placement algorithms, multimedia ASIC and hardware design.}

\profile{Song Chen}{ received the B.S. degree in computer science
from Xi’an Jiao-tong University, China, in 2000, the M.S. and Ph.D.
degrees in computer science from Tsinghua University, China, in 2003
and 2005, respectively. From August 2005 to April 2009, he had been
a visiting associate at the Graduate School of IPS, Waseda
University, Japan, where he is now an assistant professor. His
research interests include several aspects of electronic design
automation, e.g., floorplanning, placement, high-level synthesis.}

\profile{Satoshi GOTO}{ received the B.E. and M.E. degree in
Electronics and Communication Engineering from Waseda University in
1968 and 1970, respectively. He also received the Dr. of Engineering
from Waseda University in 1981. He is IEEE fellow, Member of Academy
Engineering Society of Japan and professor of Waseda University. His
research interests include LSI System and Multimedia System.}

\end{document}